\documentclass{iaus}
\usepackage{graphicx}

\title[Transiting planet parameters] 
{Toward a homogeneous set of transiting planet parameters}

\author[Torres et al.]   
{Guillermo Torres$^1$,
Joshua N. Winn$^2$,
\and Matthew J. Holman$^1$}

\affiliation{$^1$Harvard-Smithsonian Center for Astrophysics, \\
60 Garden St., Cambridge, MA 02138, USA
 \\ email: {\tt gtorres@cfa.harvard.edu, mholman@cfa.harvard.edu} \\[\affilskip]
$^2$Dept. of Physics, and Kavli Institute for Astrophysics and
Space Research, \\ Massachusetts Institute of Technology, Cambridge, MA 02139, USA
 \\email: {\tt jwinn@mit.edu}}

\pubyear{2008}
\volume{xxx}  
\pagerange{119--126}
\jname{Title of your IAU Symposium}
\editors{A.C. Editor, B.D. Editor \& C.E. Editor, eds.}
\begin{document}

\maketitle

\begin{abstract}
With 40 or more transiting exoplanets now known, the time is ripe to
seek patterns and correlations among their observed properties, which
may give important insights into planet formation, structure, and
evolution. This task is made difficult by the widely different
methodologies that have been applied to measure their properties in
individual cases. Furthermore, in many systems our knowledge of the
planet properties is limited by the knowledge of the properties of the
parent stars. To address these difficulties we have undertaken the
first comprehensive analysis of the data for 23 transiting planets
using a uniform methodology. We revisit several of the recently
proposed correlations, and find new ones involving the metallicity of
the parent stars.
\end{abstract}

\firstsection 

\section{Introduction}

Most of our knowledge about the structure, atmospheric properties, and
other physical characteristics of extrasolar planets has come from the
study of those that transit their parent stars. The pace of discovery
of transiting planets has increased rapidly over the last year or so,
and at the time of this writing there are nearly 40 systems with
detailed studies in the literature, along with several more that have
been announced recently.  The time is ripe to seek patterns and
correlations among their observed properties, which may give important
insights into planet formation, structure, and evolution.

Several such relations have already been proposed. Unfortunately, our
ability to gauge their reliability or to find new ones is made
difficult by the widely different methodologies that have been applied
by individual investigators to measure the properties of the planets
and their parent stars. Furthermore, in many cases our knowledge of
the planet properties is limited by the knowledge of the properties of
the stars themselves, as surprising as this may seem. The latter
properties are usually determined with the help of stellar evolution
models, but not always have the best constraints been applied
consistently. In particular, for the majority of transiting systems
without a parallax determination, the weakly constrained surface
gravity of the star determined spectroscopically has often been used
as a proxy for luminosity. A much better constraint related to the
mean stellar density is available directly from the light curves
(\cite[Sozzetti et al.\ 2007]{Sozzetti:07}), but has generally been
overlooked.

To address these difficulties we have undertaken the first
comprehensive analysis of the data for 23 transiting planets using a
uniform methodology (\cite[Torres et al.\ 2008]{Torres:08}). We
describe our procedures here, along with a few highlights of our
findings.

\section{Methodology}

Our efforts to re-analyze the data for all transiting planets are
focused on three main areas:

\noindent{\it Stellar atmosphere parameters:} We have merged all
existing determinations of the stellar temperature ($T_{\rm eff}$) and
metallicity ([Fe/H]) of the parent stars, with careful consideration
of systematic errors in computing the weighted averages. These
represent the best available values for these stars based on current
knowledge.

\noindent{\it Light curves:} The highest-quality light curves
available to us for each system have been re-analyzed in a uniform
way, using the Markov Chain Monte Carlo algorithm. We have accounted
for the effects of red noise in weighting the data, to provide more
realistic uncertainties for the three light-curve parameters. These
are $R_p/R_\star$ (the planet-to-star radius ratio), $b$ (the impact
parameter), and $a/R_{\star}$ (the normalized separation), where $a$
is the semimajor axis of the orbit.

\noindent{\it Stellar parameters:} We used stellar evolution models
from the Yonsei-Yale series (\cite[Yi et al.\ 2001]{Yi:01}) to
determine the mass ($M_\star$) and radius ($R_\star$) of all parent
stars in a uniform way. The constraints we used to place the stars on
the H-R diagram are $T_{\rm eff}$, [Fe/H], and the mean stellar
density ($\rho_\star$).  The density is related to the light-curve
parameter $a/R_{\star}$ as
\begin{equation}
\rho_\star = {3\pi\over G P^2}\left({a\over R_\star}\right)^3 - \rho_p \left({R_p\over R_\star}\right)^3,
\end{equation}
where $P$ is the orbital period and the second term on the right is
typically negligible.  The resulting masses and radii were checked
against those from two other sets of stellar evolution models
(\cite[Girardi et al.\ 2000]{Girardi:00}; \cite[Baraffe et al.\
1998]{Baraffe:08}), and were found to be in excellent agreement.

\section{Results}

From the reanalysis of the data for 23 transiting systems we have
obtained a more homogeneous set of stellar and planetary parameters
than previously available, with error bars that are well understood
and more appropriate when searching for patterns and correlations
among the various quantities. With these results we have revisited
several of the recently proposed correlations, of which we illustrate
two here.

\begin{figure}[h]
\begin{center}
 \includegraphics[width=3.0in]{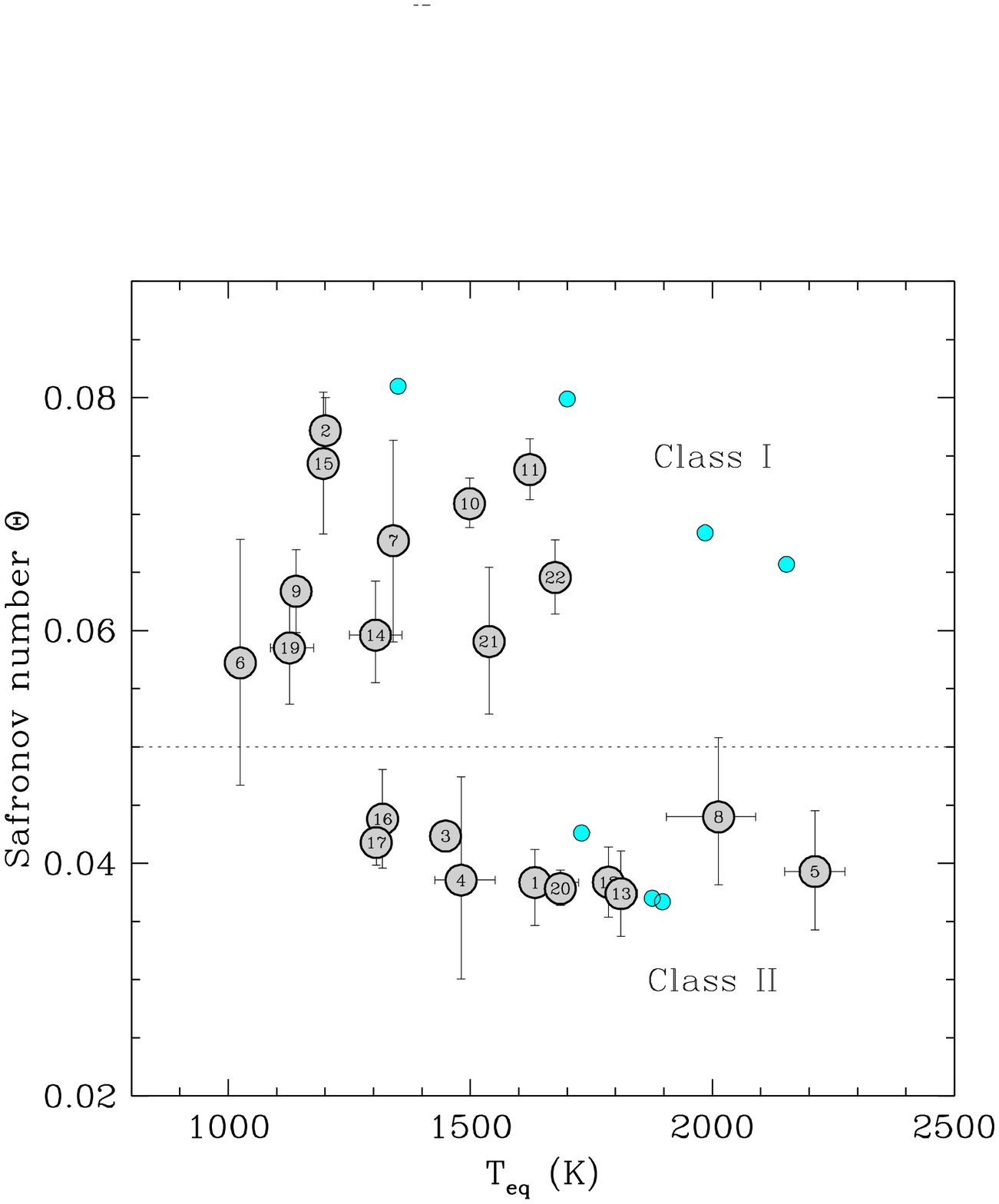}
 \caption{Safronov number versus equilibrium temperature.  HAT-P-2 and
GJ~436 are off the scale. The numbering follows that of
Figure~\ref{fig3} below. Smaller symbols represent the location of new
planets discovered since our study, which confirm the trend.}
\label{fig1}
\end{center}
\end{figure}

One is an apparent dichotomy in the properties of transiting planets
according to their Safronov numbers, advanced by \cite[Hansen \&
Barman (2007)]{Hansen:07}.  The Safronov number
is a measure of the ability of a planet to
gravitationally scatter other bodies, and is defined as $\Theta =
\frac{1}{2}(V_{\rm esc}/V_{\rm orb})^2 = (a/ R_p)(M_p/ M_{\star})$,
the ratio between the escape velocity and the orbital velocity
squared.  Figure~\ref{fig1} shows $\Theta$ as a function of each
planet's zero-albedo equilibrium temperature, which we compute as
$T_{\rm eq} = T_{\rm eff}(R_{\star}/2a)^{1/2}$ (assuming that the heat
redistribution factor is common to all planets, in the absence of more
complete knowledge).  The conspicuous gap between ``Class~I'' and
``Class~II'' planets noted by Hansen \& Barman is reinforced with the
addition of the new planets in our study, and the clustering of the
Class~II planets is tightened. We also find that Class~II planets
orbit stars that are slightly more metal rich than those of Class~I,
on average, by $\sim$0.2 dex in [Fe/H].

\begin{figure}[h]
\begin{center}
 \includegraphics[width=5.3in]{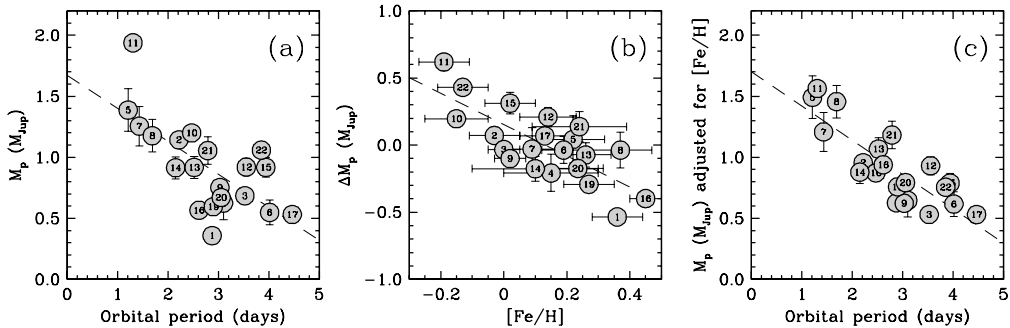}
 \caption{(a) $M_p$ versus orbital period for all transiting planets
in our sample except HAT-P-2 and GJ~436. (b) $O\!-\!C$ residuals
$\Delta M_p$ from the linear fit in the left panel, shown as a
function of the metallicity of the host star. (c) Same as (a), with
the [Fe/H] dependence removed.}  \label{fig2}
\end{center}
\end{figure}

\cite[Mazeh et al.\ (2005)]{Mazeh:05} and \cite[Gaudi et al.\
(2005)]{Gaudi:05} pointed out a correlation between the masses of
transiting planets and their orbital periods.  This relation has held
up as more transiting planets have been discovered, although the
scatter has increased. We have found evidence for a new pattern within
the scatter about this correlation: planets around metal-poor stars
are more massive than those around metal-rich stars at a given orbital
period (see Figure~\ref{fig2}). This can be interpreted as evidence of
a (mass and) metallicity dependence of the migration process.
Alternatively, it may be seen as indirect support for the correlation
between core size and [Fe/H] (\cite[Guillot et al.\ 2006]{Guillot:06};
\cite[Burrows et al.\ 2007]{Burrows:07}) along with the idea that the
presence of such cores slows down or prevents complete evaporation of
the planets in the extreme radiation environments of these hot
Jupiters. A related correlation between the planetary surface gravity
(a quantity that is virtually independent of the stellar mass and
radius) and orbital period has been pointed out previously by
\cite[Southworth et al.\ (2007)]{Southworth:07}. We find that this
correlation is also present in our larger sample, but the scatter
about the relation does not show as clear a dependence on metallicity,
possibly because of the influence of [Fe/H] on the planetary radii.

The figure shown below is a visual summary of our derived stellar and
planetary properties. The period of each system is indicated. Stars
and planets are shown to scale, emphasizing the wide range of stellar
types probed by the photometric searches. The equatorial plane is
indicated with a solid horizontal line, and the dotted lines represent
the trajectory of each planet at their measured impact parameter. The
light curves underneath are computed for the $V$ band, and are all on
the same vertical and horizontal scale to facilitate the comparison.

\begin{figure}[h]
\begin{center}
 \includegraphics[width=5.2in]{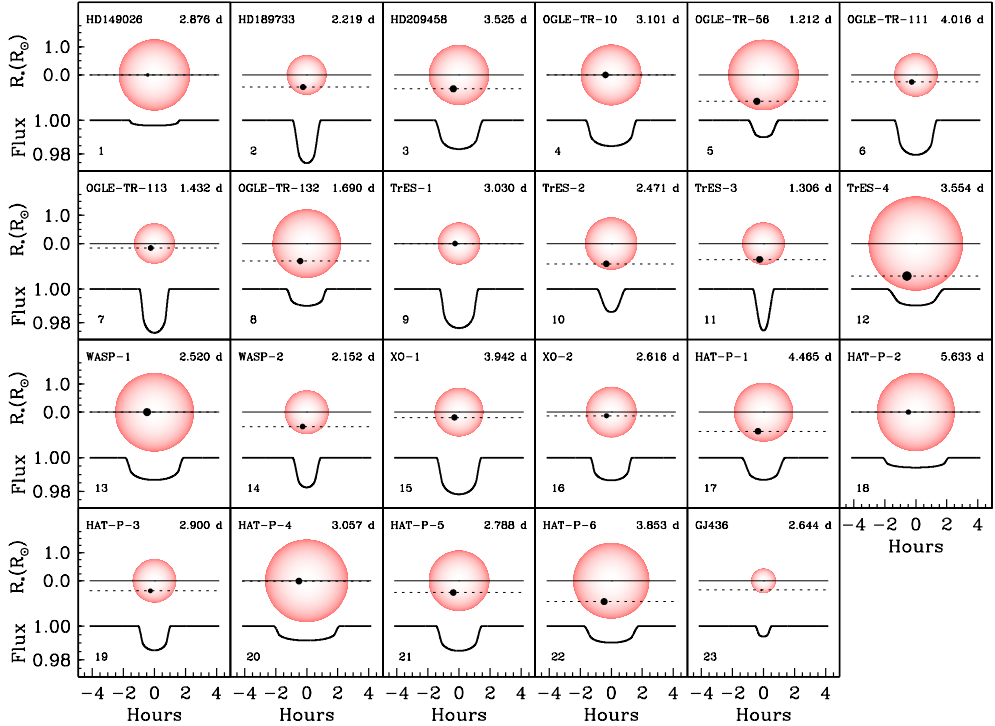}
 \caption{``Portrait gallery'' of the 23 transiting planets in our study.}
   \label{fig3}
\end{center}
\end{figure}

Through the procedures described above we have obtained a more
homogeneous set of stellar and planetary parameters than previously
available. We expect that the application of similar methods to future
discoveries will make it easier to search for other significant
correlations among those parameters that should lead to a deeper
understanding of the nature of extrasolar planets.

\end{document}